\documentclass{aa}  

\usepackage[dvipsnames]{xcolor}
\usepackage{graphicx}
\usepackage{txfonts}
\usepackage{hyperref}
\usepackage{float}
\hypersetup{
    colorlinks=true,
    linkcolor=blue,
    citecolor=blue,
    filecolor=magenta,      
    urlcolor=cyan,
    pdftitle={Overleaf Example},
    pdfpagemode=FullScreen,
    }
\usepackage{graphicx}
\usepackage{subcaption}

\newcommand{\jwst}{{\it JWST}}

\begin{document} 
    \title{Too many protoclusters? Reconciling the overabundance of cluster progenitors within the first billion years of the Universe}
    \titlerunning{Reconciling the overabundance of massive protoclusters within the first billion years}

    \author{Callum Witten\inst{1}
    \and Jake S. Bennett\inst{2}
    \and Pascal A. Oesch\inst{1,3,4}
    \and Seunghwan Lim\inst{5,6}
    \and Chamilla Terp\inst{3,4}
    \and Jakob M. Helton\inst{7}
    \and Kasper E. Heintz\inst{8,3,4}
    \and Romain A. Meyer\inst{1}
    \and William McClymont\inst{5,6}
    \and Thomas Herard-Demanche\inst{9}
    \and Emma Giovinazzo\inst{1}
    }

   \institute{Department of Astronomy, University of Geneva, Chemin Pegasi 51, 1290 Versoix, Switzerland
   \and
   School of Physics \& Astronomy, University of Nottingham, University Park, Nottingham, NG7 2RD, UK
   \and
   Cosmic Dawn Center (DAWN), Copenhagen, Denmark
   \and
   Niels Bohr Institute, University of Copenhagen, Jagtvej 128, 2200 Copenhagen, Denmark
    \and
    Kavli Institute for Cosmology, University of Cambridge, Madingley Road, Cambridge, CB3 0HA, UK
    \and
    Cavendish Laboratory, University of Cambridge, 19 JJ Thomson Avenue, Cambridge, CB3 0HE, UK
   \and
   Department of Astronomy and Astrophysics, The Pennsylvania State University, University Park, PA 16802, USA
    \and
    DTU Space, Technical University of Denmark, Elektrovej 327, DK2800 Kgs. Lyngby, Denmark;
    \and
    Leiden Observatory, Leiden University, PO Box 9513, 2300 RA Leiden, The Netherlands 
    \\
   \email{callum.witten@unige.ch}
             }

   \date{Submitted 26 May 2026}

  \abstract
    {\jwst\ has revealed an overabundance of apparent (`Coma'-like) cluster progenitors at $z>5$ that is $\sim$ two (300) times in excess of, and in $2.3\sigma$ ($4\sigma$) tension with, the number density of such objects at present day and theoretical predictions from $\Lambda$CDM. We present an analysis of protocluster candidates between $5<z<9$ from the literature, and in the TNG-Cluster and TNG300 simulations, aimed at resolving this tension. We first identify an inconsistency in how protocluster candidates are characterised: observational halo masses are estimated by summing the stellar mass over very large apertures ($\gtrsim 50\, R_{\rm vir}$), but are compared to the halo mass within the virial radius from simulations. In addition, these halo masses are commonly compared to the backward-tracked halo mass evolution of clusters, whereas a forward-tracked evolution of massive high-redshift haloes is more appropriate. Correcting these inconsistencies, while accounting for the merging of coincident protoclusters and the fluctuations associated with cosmic variance, entirely alleviates the tension. Ultimately, we find that $64\%$ of the protocluster candidates are instead likely to be proto-groups (i.e., $M_{200c}<10^{14}\, {\rm M_{\odot}}$) and none appear likely to become Coma-like clusters (i.e., $M_{200c}>10^{15}\, {\rm M_{\odot}}$) by $z=0$. Although some of these overdensities may not become clusters, they still trace extreme nodes of the cosmic web that may host early environmental effects, drive the first ionised regions, and contribute significantly to the cosmic star-formation rate density. These results demonstrate the need for more careful comparisons between observations and simulations of high-redshift protoclusters, and that improved selection criteria, potentially using the summation of the halo mass on Lagrangian scales, are vital for high-redshift protocluster science. 
    }
   \keywords{galaxies: high-redshift -- galaxies: clusters: general --  dark ages, reionization, first stars -- large-scale structure of Universe}
   \maketitle

\section{Introduction}
\label{sec:intro}
The formation of structure throughout the Universe is remarkably well described by the $\Lambda$ cold dark matter ($\Lambda$CDM) model \citep{Peebles1982, Blumenthal1984}. While small-scale perturbations in the dark matter distribution result in the formation of galaxies in the early Universe, large-scale overdensities, with time, collapse into massive clusters of galaxies. At present day we define galaxy clusters to be dark matter haloes with masses of $M_{\rm halo}>10^{14}\,{\rm M_{\odot}}$ \citep[e.g.][]{Voit2005, Overzier2016}. The progenitors of these, coined protoclusters \citep{Overzier2016}, are thought to be some of the most massive haloes in the early Universe hosting extreme overdensities of galaxies \citep{Chiang2013}. 

At low redshift ($z<2$) it is possible to measure an accurate halo mass using X-ray emission, the Sunyaev–Zeldovich (SZ) effect and velocity dispersion measurements \citep[e.g.][]{Stanford2012,Brodwin2012,Newman2014,Mantz2014,Bleem2015}. However, in the first billion years, this is increasingly challenging. Protoclusters are in the early stages of virial heating and the development of a hot circumgalactic medium (see \citealt{Bennett2026} for further discussion), meaning X-ray and SZ detections are very difficult. While detections have been made at high redshift \citep{Zhou2026,Bogdan2026}, these are often tentative and may produce biased halo mass measurements, due to higher levels of non-thermal pressure support in unvirialised haloes \citep[see e.g.][]{Bennett2022}. Moreover, the galaxies resident within high-redshift protoclusters are also unlikely to be virialised, and as such, their velocity dispersions will not trace the virial mass of the structure. Therefore, at high redshift, while galaxy overdensities are typically assumed to trace dark matter overdensities, and hence are thought to be key to identifying potential protocluster candidates \citep[e.g.][]{Chiang2013}, alternative indirect tracers of their halo masses are required.

The sensitivity and spectral coverage of \jwst\ has facilitated the detection of an abundance of galaxy overdensities at $z>5$ \citep{Laporte2022, Hashimoto2023, Morishita2023,Morishita2025, Witten2025, Fudamoto2025, Helton2024, Helton2024a, Scholtz2024, Castellano2023, Kashino2023, Meyer2024, Meyer2025, Terp2026, Li2025, schindler_little_2025, Bogdan2026, HerardDemanche2025,Wu2025, Witstok2024, Whitler2024,Champagne2025, Wu2026, Witten2026}. These overdense regions of the early Universe are seen to produce some of the most evolved galaxies at high redshift \citep{Arribas2024,Morishita2025, Witten2025b,Umehata2025, Fudamoto2025a, Witten2026}, although this is potentially a function of halo mass (see \citealt{Witten2025, Gelli2025, Terp2026} for further discussion). They are also thought to be the engines behind the first ionised regions \citep{Lu2024, Witstok2024, Chen2026, Martin2026}. 

These overdensities range from large-scale collections of relatively low-mass galaxies \citep[e.g.][]{Fudamoto2025} to small-scale clusters of massive objects \citep[e.g.][]{Hashimoto2023, Fudamoto2025a, Wei2026}, and it is therefore important to consider which environments trace the most massive dark matter haloes. Previous works have often focused on summing the halo masses of all galaxies within the overdensity to establish the `total' halo mass. These total halo masses reported in the literature are typically summed over very large apertures. These have sizes that are comparable to the protocluster Lagrangian radius ($R_{\rm L}\sim 10-15$ cMpc at $z>5$; \citealt{Chiang2013}), which is defined to be the radius encapsulating all galaxies that reside within the eventual cluster at $z=0$.

Often, this total halo mass is then compared to the typical halo mass of `Coma'-like cluster ($M_{\rm halo}>10^{15}\,{\rm M_{\odot}}$; \citealt{Zwicky1933,Zwicky1937}) progenitors at a given redshift from the simulation analysis of \cite{Chiang2013}. \cite{Chiang2013} utilise the Millennium Run N-body simulation \citep{Springel2005} with semi-analytic galaxy formation models from \cite{Guo2011} to follow the most massive progenitors, i.e. the backward-tracked evolution, of present-day Coma-like clusters (see Figure 2 of \citealt{Chiang2013}). Such simulations measure the halo mass ($M_{200c}$) within $R_{200c}$, which is defined as the radius of the volume within which the mass density is 200 times the critical density of the Universe, $\rho_{\rm crit}(z)$, at a given redshift:
\begin{equation}
\label{eq:R200}
    R_{200c} = \left[\frac{3M_{200c}}{4\pi\times 200\, \rho_{\rm crit}(z)}\right]^{1/3}\, .
\end{equation}
For halo masses ranging from $M_{200c}\, [{\rm M_{\odot}}]= 10^{11}-10^{12}$, the associated virial radius ranges from $R_{200c}\sim 15-50$ pkpc in the redshift range $5<z<9$. 

When the observed total halo mass is found to be consistent with, or in excess of, the $M_{200c}$ evolution from \cite{Chiang2013}, these overdensities are typically reported to be protocluster candidates. While this is assumed to be an effective method for selecting protoclusters, a notable challenge arises: a significant excess in the number density of protocluster candidates relative to $\Lambda$CDM predictions. This excess has been previously noted by \cite{Helton2024}, \cite{Lim2024} and \cite{Fudamoto2025}, and even for comedic effect by \cite{Lovell2025}. This discrepancy is relative to the cluster mass function at present day, which is well constrained by a combination of simulations \citep{Bahcall1993,Jenkins2001,Evrard2002,Warren2006,Tinker2008, Angulo2012} and observations \citep{Girardi1998,Reiprich2002,Vikhlinin2009,Bohringer2017,Abdullah2020, Hung2021}; finding the number density of clusters (i.e., $M_{200c}>10^{14}\,{\rm M_{\odot}}$) to be $n\sim 1\times 10^{-5}\, {\rm cMpc^{-3}}$. 

The discrepant expected and observed number densities of protoclusters can be a result of two incorrect assumptions: (i) a single present-day cluster is comprised of a single high-redshift protocluster, i.e. protoclusters do not merge, and, (ii) that we have an accurate and robust selection criterion for protocluster environments. While the first scenario can help to reduce the number density of distinct cluster progenitors \citep[e.g.][]{Helton2024}, the often large redshift separations between protocluster candidates makes merging highly unlikely in other studies \citep[such as][]{Terp2026}. Therefore, in order to explain their excessive abundance, the sample of literature protocluster candidates must be contaminated with regions of the Universe that are not on-course to become clusters at present day. This implies that either the total halo mass estimate, or the application of the halo mass evolutionary pathway from \cite{Chiang2013}, is not appropriate. 

An alternative solution would be that the halo mass evolutionary pathways from simulations are not representative of reality due to faults in our assumed cosmological model, $\Lambda$CDM. For example, the Early Dark Energy (EDE) model \citep{Karwal2016, Poulin2018, Poulin2019, Smith2020} was proposed as a solution to the excessive number density of UV-bright galaxies seen in the high-redshift Universe by \cite{Shen2024}. This overshoots the $\Lambda$CDM halo mass function at all halo masses at high redshift, while converging with the $\Lambda$CDM halo mass function by $z<2$. As such, changing cosmological models to EDE could feasibly overproduce the number of massive dark matter haloes in the early Universe, while resulting in the well constrained cluster number density by today. 

The excess number density of massive haloes could have serious implications for galaxy evolution, potentially providing a solution to the rapid appearance of mature galaxies in overdensities at high redshift \citep[e.g.][]{Helton2024a, Fudamoto2025a, Li2025, Witten2026}. Additionally, this could lead to an early onset to reionisation, potentially resulting in the formation of large ionised regions around overdensities of high-redshift galaxies \citep[e.g.][]{Witten2024,Scholtz2024,Lu2024,Tang2024,Witstok2025, Chen2026}. It is therefore crucial to understand what drives this theoretical tension, and whether it can be resolved.

While this paper focuses on the use of the halo mass of overdensities to diagnose protoclusters, we also note that some previous works have attempted to utilise an overdensity parameter threshold, also proposed by \cite{Chiang2013}, to identify protocluster candidates \citep[e.g.][]{Helton2024}. However, this suffers from similar challenges to the aforementioned method, including a diversity of protocluster evolutionary phases at a given redshift, and that the progenitors of clusters and groups can host similar galaxy overdensities (see \citealt{Chiang2013} for further discussion). As a result, in this paper, we focus on the method of identifying a candidate massive halo at high redshift, and using halo mass evolutionary pathways to determine whether it is a protocluster, especially given that this is the most commonly used high-redshift protocluster selection criteria in the literature \citep{Long2020,Laporte2022,Li2025,Helton2024a,Terp2026}.

We first present our sample of $z\sim5-9$ protocluster candidates taken from the literature in Section~\ref{sec:sample} and infer the number density of protoclusters in Section~\ref{sec:nominal_number_density}. We attempt to resolve the theoretical tension in Section~\ref{sec:resolution} and discuss the broader implications of our findings in Section~\ref{sec:implications}, before concluding in Section~\ref{sec:conclusion}.

In order to understand whether the tension in the number density of protoclusters can be resolved in a $\Lambda$CDM universe, throughout this paper, we adopt a standard $\Lambda$CDM cosmology, assuming the cosmological parameters from \cite{Collaboration2020}. We additionally highlight that we use the term virial radius throughout to indicate $R_{200c}$. While the virial radius and $R_{200c}$ are not identical, they are very similar at high redshift as the Universe is matter-dominated \citep{Peebles1980,Bryan1998}, and thus we use these terms interchangeably. 

\section{Sample of protoclusters}
\label{sec:sample}

Within this work we utilise existing samples of protoclusters identified from searches in deep \jwst\ NIRCam imaging or grism spectroscopy. These include those identified by \cite{Terp2026} in the Abell 2744 field and by \cite{Helton2024} in the GOODS-South and GOODS-North field, both with grism spectroscopy, and by \cite{Li2025} in the EGS and NEP fields with imaging. All of these protocluster candidates are reported in Tables~\ref{tab:spec_sample} and ~\ref{tab:phot_sample}.

While each work employs a slightly different method, generally, these works target overdensities of galaxies relative to the expected number density, and estimate a halo mass for each overdensity. The halo masses of these protoclusters are computed in a broadly homogenous manner\footnote{While \cite{Terp2026} employ a range of different methods for estimating the halo mass, the stellar-to-halo mass conversion method typically offers the lowest halo mass of these methods, which is the most favourable scenario to produce the lowest protocluster number density.}: by converting the observed stellar mass of overdensity-resident galaxies to a halo mass estimate, using a redshift-dependent stellar mass-halo mass (SMHM) relation, e.g. \citep{Behroozi2019}. While different studies do not employ a consistent SMHM relation, different relations produce minimal changes in the inferred halo mass \citep[e.g. $\Delta M_{\rm halo}<0.2\, {\rm dex}$;][]{Girelli2020}. They then compare this halo mass to the typical halo mass evolution of a galaxy cluster \citep[e.g.][]{Chiang2013}, and report that these overdensities likely represent the progenitors of present-day clusters, i.e. protoclusters. 

\begin{figure*}
    \centering
    \includegraphics[width=1\linewidth]{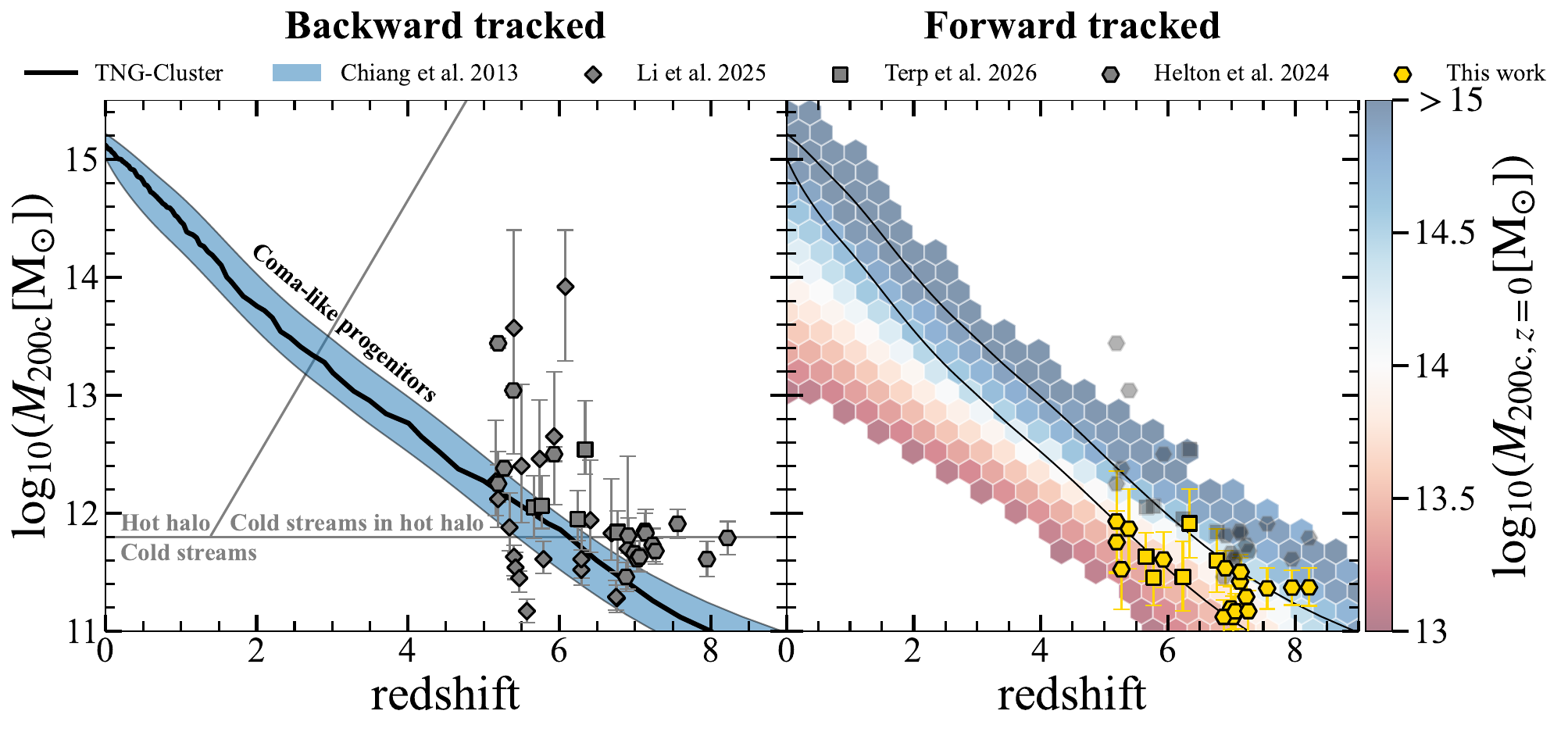}
    \caption{{\bf a:} (left) Backward-tracked evolution of the halo mass of Coma-like clusters from \cite{Chiang2013} (blue shaded region) and from the TNG-Cluster simulations (black line). Over-plotted are the literature halo masses of protocluster candidates from \cite{Li2025} (grey diamonds), \cite{Terp2026} (grey squares) and \cite{Helton2024} (grey hexagons). The grey lines denote the typical phases of circumgalactic medium gas from \cite{Dekel2006}. {\bf b:} (right) Forward-tracked evolution of massive high-redshift haloes in the TNG-Cluster and TNG300 simulations. The colour of each hexbin indicates the median present-day halo mass (when it is above $10^{13}\, {\rm M_{\odot}}$) of haloes that fall within said hexbin. The transparent grey data points indicate the literature halo masses, while the yellow data points indicate the halo masses as a result of the adapted SMHM relation derived in this work (the symbols are as previously defined). The black lines indicate the envelope of the \cite{Chiang2013} Coma-like cluster backward-tracked halo mass evolution, allowing for comparison between the forward tracking method implemented in this work. Combining our adapted SMHM relation and forward-tracked halo mass evolution, it becomes apparent that previously reported proto-Coma-clusters are instead likely to become lower mass haloes (i.e. $M_{200c}<10^{15}\, {\rm M_{\odot}}$) by present day.}
    \label{fig:M200_evolution}
    \phantomsubcaption\label{fig:M200_evolution_a}
    \phantomsubcaption\label{fig:M200_evolution_b}
\end{figure*}

\begin{figure}
    \centering
    \includegraphics[width=1\linewidth]{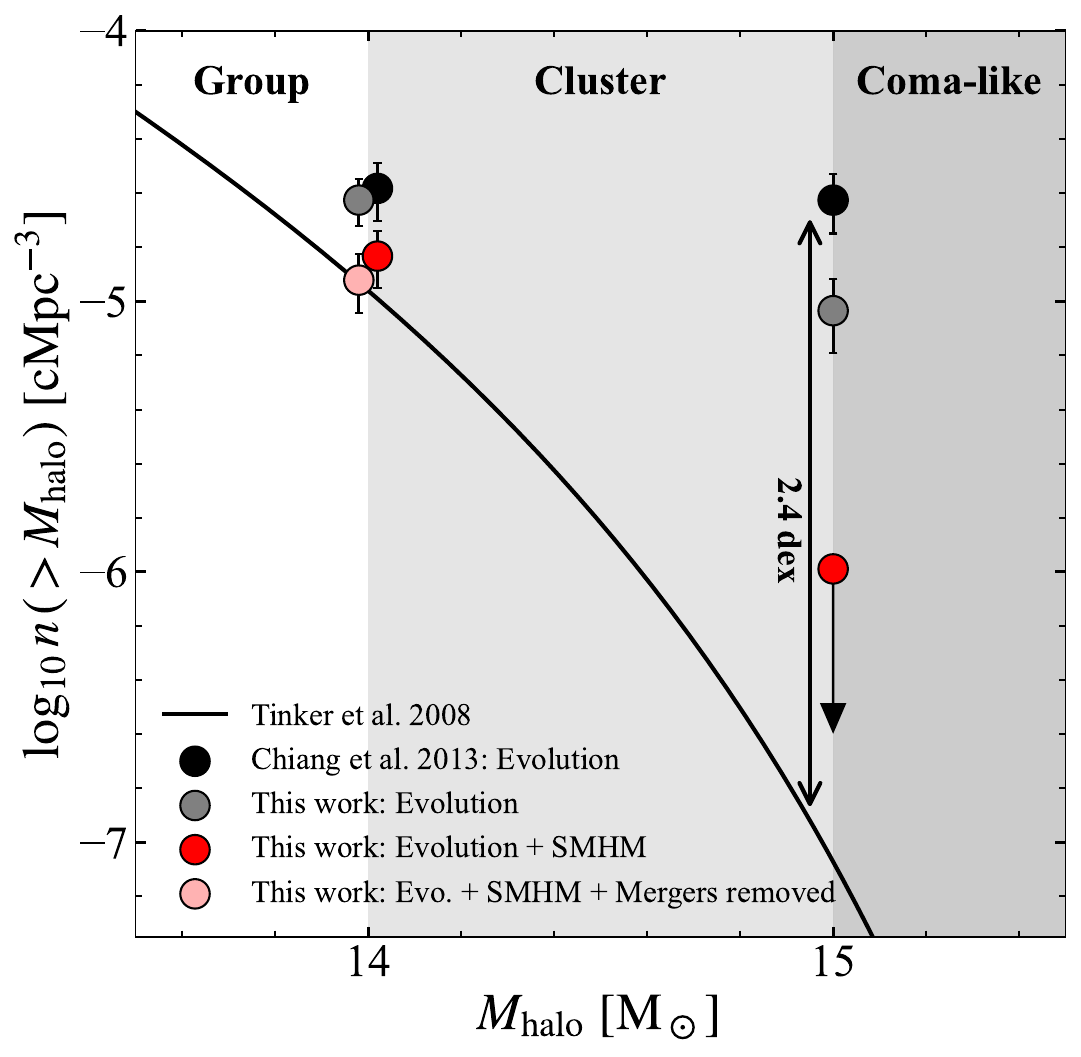}
    \caption{Number density of protocluster candidates: reported in the literature using the \cite{Chiang2013} backward-tracked halo mass evolution (black), when using our forward tracking halo mass evolution to identify protoclusters (grey) and when combined with re-estimating their halo mass with our adapted SMHM relation (red). The transparent red data point indicates the number density when we account for the merging of coincident protocluster candidates. The number densities of protoclusters are offset by 0.02 dex on the x-axis for ease of viewing. The black line indicates the halo mass function at $z=0$ from \cite{Tinker2008}. The number density of (Coma-like) cluster progenitors should be equal to the number density of (Coma-like) clusters at $z=0$, however, the nominal number density from literature samples is significantly greater than this -- in $2.3\sigma$ ($4\sigma$) tension.}
    \label{fig:number_density}
\end{figure}

In order to estimate the number density of protoclusters, we need to evaluate the volume of each survey. The protocluster searches from \cite{Li2025}\footnote{We exclude their protocluster search in GOODS-S which is already covered by the spectroscopic search of \cite{Helton2024}.} and \cite{Helton2024} are conducted in non-lensed fields, and as such, using the surface areas (158 arcmin$^2$ and 81 arcmin$^2$, respectively) and redshift ranges ($5<z<7$ and $4.9<z<8.9$, respectively) we estimate the search volumes to both be $7.6 \times 10^5 \, {\rm cMpc^3}$. In contrast, the \cite{Terp2026} analysis occurs in the Abell 2744 cluster field, and the resulting lensing complicates volume estimates. As such, we simply take the volume reported by \cite{Terp2026}, resulting in a total volume of $8.36 \times 10^4 \, {\rm cMpc^3}$. The volumes of each survey, as well as the number of protocluster candidates, $N_{\rm PC}$, are reported in Table~\ref{tab:density}.

These volumes are explicitly the total volume within which each work searches for protocluster candidates. However, a number of factors, including edge-of-field effects, redshift-dependent selection effects and filter transmission functions (for example, see \citealt{Meyer2025}) ultimately means the effective volume is slightly smaller than the total search volume.

\section{The number density of $z\sim5-9$ protoclusters in the early Universe}
\label{sec:nominal_number_density}

\begin{table}[]
    \centering
    \caption{Number density of protoclusters inferred from literature searches.}
    \label{tab:density}
    \begin{tabular}{lccc}
    \hline
    Survey & Volume $[{\rm cMpc^{3}}]$ & $N_{\rm PC}$ & $n\, [{\rm cMpc^{-3}}]$\\ 
    \hline
    {\it Spectroscopic} \\
    \cite{Terp2026} & $8.4 \times 10^4$ & 5 & $6.0 \times 10^{-5}$\\
    \cite{Helton2024} & $7.6 \times 10^5$ & 17 & $2.2 \times 10^{-5}$ \\
    \hline
    {\it Photometric} \\
    \cite{Li2025} & $7.6 \times 10^5$ & 20 & $2.6 \times 10^{-5}$ \\
    \hline
    {\bf Total} & $1.6 \times 10^6$ & 42 & $2.6 \times 10^{-5}$ \\
    \hline
    \end{tabular}
\end{table}

Present-day galaxy clusters are defined as having a halo mass of $M_{\rm 200c}\geq10^{14}\,{\rm M_{\odot}}$, and hence typical searches for protocluster candidates have focussed on identifying the most massive dark matter haloes at a given redshift. The stellar mass of a galaxy is related to the mass of its dark matter halo, and hence overdensities of massive galaxies are thought to trace the most massive dark matter haloes. 

The key determinant of whether a high-redshift massive halo will become a cluster at present-day is whether the inferred halo mass lies on, or above, the halo mass evolution of a galaxy cluster with redshift. In order to assess this, the halo masses of high-redshift protocluster candidates are typically compared to the evolutionary pathway of Coma-like clusters, often from \cite{Chiang2013} \citep[e.g.,][]{Long2020,Laporte2022,Li2025,Helton2024a,Terp2026}. The \cite{Chiang2013} halo mass evolution is shown in Figure~\ref{fig:M200_evolution_a} compared to the reported halo masses of our sample of protoclusters; almost all of our sample lie on or above this evolution of a Coma-like cluster. This suggests that all but four of our sample will become massive present-day clusters with halo masses of $M_{\rm 200c}\geq10^{15}\,{\rm M_{\odot}}$. This therefore results in a number density of Coma-like cluster progenitors (hereafter proto-Coma-clusters) of $n=2.36 \times 10^{-5} \, {\rm cMpc^{-3}}$. Given the proximity of the remaining four halo masses to the \cite{Chiang2013} Coma-like cluster evolution, we assume that the remaining four are on course to become regular clusters ($M_{\rm 200c}\geq10^{14}\,{\rm M_{\odot}}$), and hence, that all 42 of the sample are protocluster candidates. We therefore find the number density of protoclusters to be $n=2.6 \times 10^{-5} \, {\rm cMpc^{-3}}$, and note that the four additional protoclusters that we include have a minimal impact on this value. The number density of protocluster candidates in each survey are reported in Table~\ref{tab:density}.

The present-day halo mass function is well constrained \citep[e.g.][]{Tinker2008}, and as a result the number density of clusters, and indeed Coma-like clusters, are known to be $n_{M_{\rm halo}>10^{14}\, {\rm M_{\odot}}}\sim 1 \times 10^{-5} \, {\rm cMpc^{-3}}$ and $n_{M_{\rm halo}>10^{15}\, {\rm M_{\odot}}}\sim 1 \times 10^{-7} \, {\rm cMpc^{-3}}$, respectively. We can compare the present-day cluster number density to the high-redshift protocluster number density if we assume that multiple cluster progenitors do not merge to form a single more massive cluster. We discuss how this may affect the inferred number density of protoclusters further in Section~\ref{subsec:merging}. In the following, however, we simply assume each protocluster forms a distinct cluster, and hence the number density of protoclusters should equal the number density of present-day clusters. 

In Figure~\ref{fig:number_density} we show that the number density of proto-Coma-cluster candidates reported in the literature is 2.4 dex higher than the number density of present-day Coma-like clusters predicted by \cite{Tinker2008}. This discrepancy lies far beyond the uncertainties driven by the Poisson noise, or indeed cosmic variance (see discussion below), suggesting that we observe far too many proto-Coma-cluster structures in the early Universe. If we instead assume that all of the objects in Tables~\ref{tab:spec_sample} and \ref{tab:phot_sample} are simply cluster progenitors, i.e. $M_{{\rm halo},\, z=0}>10^{14}\, {\rm M_{\odot}}$,
the discrepancy remains. As has previously been noted by \cite{Helton2024} and \cite{Li2025}, we find the observed number density to be enhanced by 0.3 dex relative to the present-day number density of galaxy clusters.

In order to estimate the uncertainty in these number densities we include both Poisson noise and cosmic variance. For the latter, we estimate the dark matter variance of the survey volume ($8.4\times10^5\, {\rm cMpc^3}$) to be $\sigma_V\sim 0.02$ using {\it QuickCV} \citep{Newman2002}. The typical halo bias of haloes with masses similar to our protocluster sample ($M_{\rm halo}=10^{11}-10^{12}\, {\rm M_{\odot}}$) at $z=5-9$ ranges from $b \sim 4-25$ \citep{Tinker2010,Bhattacharya2011}. This corresponds to number density fluctuations of $\sigma_{\rm CV}=b\,\sigma_V = 0.07-0.43$. If we assume the median redshift and halo mass from Tables~\ref{tab:spec_sample} and ~\ref{tab:phot_sample}, we find $\sigma_{\rm CV}\sim 0.2$. When we combine these fluctuations from cosmic variance with the Poisson noise, we obtain the uncertainties shown in Figure~\ref{fig:number_density}. These indicate that the observed number density of protoclusters in the literature is in $2.3\sigma$ tension with the number density of observed clusters at $z=0$, while the number density of proto-Coma-clusters is in $4\sigma$ tension.

\section{Solutions to the theoretical tension}
\label{sec:resolution}

In the following section we investigate a number of solutions to the overabundance of protoclusters and proto-Coma-clusters at $z>5$. 

\subsection{Do galaxy overdensities trace massive dark matter haloes?}
\label{subsec:SMHM}

As we discussed earlier, searches for protoclusters often assume a SMHM relation \citep[e.g.,][]{Behroozi2019} to convert {\it all} of the observed protocluster-resident stellar mass into a total halo mass \citep[e.g.][]{Laporte2022,Helton2024,Fudamoto2025,Terp2026}. However, it is vital to note that simulations calculate the halo mass within the virial radius and hence comparing this to the observational halo mass summed over significantly larger scales is not appropriate. The impact of these inconsistent apertures has previously been noted by \cite{Lim2024} as a potential solution to the excess number density reported in the literature. 

In more detail, simulations measure the halo mass, $M_{200c}$, within $R_{200c}$: the radius of a sphere with a mean density that is 200 times the critical density of the Universe (i.e., Equation~\ref{eq:R200}). 
Therefore, in order to map our observed halo masses onto the halo mass evolutionary pathways from simulations, it is crucial to compute the halo mass from observations in a consistent manner, i.e. within the virial radius. Moreover, protoclusters can host a number of similarly massive galaxies within their virial radii \citep[e.g.][]{Helton2024a,Arribas2024,Fudamoto2025a, Witten2026}, and thus simple SMHM relations, utilising only the mass of the `central' galaxy in the halo, may not be appropriate. It is therefore important to establish an improved process for estimating the halo mass of high-redshift protoclusters. Such a process was suggested by \cite{Witten2025}. They derived an adapted SMHM relation, which uses the sum of the stellar mass within twice the stellar half-mass radius of subhaloes whose centres fall within $R_{200c}$, as a proxy for total stellar mass within the virial radius, rather than merely the SMHM relation derived from the stellar mass of the central \citep[e.g.][]{Behroozi2019}. 

Following this work, we re-estimate the halo mass of the protocluster candidates reported by \cite{Helton2024} and \cite{Terp2026}\footnote{We utilise the spectroscopic protocluster samples as this removes chance alignments as a result of photometric uncertainty. We do note that the redshift dispersion of the spectroscopic samples is largely dominated by the relative velocity of objects and hence in the following we simply use their on-sky separation.}. We first measure the stellar mass within the virial radius of haloes from the TNG300 \citep{Pillepich2018,Nelson2018,Naiman2018,Marinacci2018,Springel2018} and TNG-Cluster \citep{Nelson2024} simulations, producing an adapted SMHM relation at $z=5,\,6,\,7,\,8\ {\rm and }\ 9$. Given that the mass resolution of TNG-Cluster and TNG300 is $\sim 10^7\, {\rm M_\odot}$ we only fit our adapted SMHM relation when $M_{\star,\,200c}>5\times 10^8\, {\rm M_\odot}$. Our adapted SMHM relations do not represent drastic offsets from the relations by \cite{Behroozi2019} (e.g., see Figure 6 of \citealt{Witten2025}), however, as we discussed before, the use of the total stellar mass within the virial radius is more applicable for these observed high density regions.

In order to apply this adapted SMHM relation to our observed sample, we need to estimate the stellar mass within the virial radius of each protocluster candidate. To do so, we perform an iterative process, given that the virial radius is defined by the halo mass (i.e., Equation~\ref{eq:R200}). We first assume a relatively large radius of $R = 100$ pkpc\footnote{This exceeds the typical virial radius of massive haloes. It is chosen to place an upper limit on the size of the halo, which is reduced iteratively, until we reach the true virial radius and mass.}, we sum the stellar mass within this distance of each galaxy in each protocluster candidate in \cite{Helton2024} and \cite{Terp2026}. We then utilise our adapted SMHM relation to estimate the halo mass within a radius of $R$ around each galaxy. With this halo mass in hand we re-estimate the virial radius, updating $R$, and recompute the total stellar mass, and hence halo mass, within said virial radius, repeating this process until convergence. This process itself is then repeated while perturbing the stellar masses of every galaxy by their associated uncertainties, to obtain the distribution of the halo mass thanks to this stellar mass uncertainty. As a result, we find the halo mass, and its associated uncertainty (including the scatter in the SMHM relation -- typically $\sim 0.05-0.15$ dex), within the virial radius, centred on each galaxy in the protocluster. The halo mass of the most massive halo in the protocluster therefore represents the $M_{200c}$ of the protocluster. Ultimately, this process simply identifies the region of each protocluster that contains the most stellar mass within its respective virial radius (which ranges from $20-50$ pkpc for our sample). These improved halo mass estimates are reported in Table~\ref{tab:spec_sample} and shown in Figure~\ref{fig:M200_evolution_b}. 

We find that all of the sample have halo masses that are inconsistent within their uncertainties with the literature values. On average, these are reduced by $\sim 0.5$ dex, and in some cases by $\sim 1.5$ dex; similar reductions were reported by \cite{Lim2024} when correcting for the large observational apertures. It is true, however, that the largest literature halo masses remain the largest haloes when we apply our adapted SMHM relation, which offers encouragement that $M_{200c}$ does likely trace the total halo mass in a much larger volume, which may be a more effective diagnostic of the present-day halo mass (see Section~\ref{sec:implications} for further discussion). When we compare these reduced halo masses to the \cite{Chiang2013} Coma-like cluster evolutionary pathway, only one of the candidates has an inconsistent halo mass, within uncertainties, with this pathway (the updated halo masses can be seen relative to the \cite{Chiang2013} pathway in Figure~\ref{fig:M200_evolution_b}). Thus, the number density of proto-Coma-clusters remains relatively unaffected.

\subsection{Are observed high-redshift massive haloes actually protoclusters?}
\label{subsec:evo}

A key component in diagnosing a protocluster candidate is inferring that it lies on a halo mass evolutionary pathway towards a present-day halo mass of $M_{200c}>10^{14}\, {\rm M_{\odot}}$. In order to do so, many previous works have compared their halo mass estimate to the evolutionary pathway of Coma-like clusters in the Millennium Simulations, reported by \cite{Chiang2013}. We first confirm that we are able to recover this halo mass evolution by following the most massive progenitor branches of the \texttt{SubLink} merger trees \citep{RodriguezGomez2015} of Coma-like clusters in the TNG-Cluster simulations. When we calculate the median halo mass of these eventual clusters at each snapshot, we find an evolutionary pathway that closely matches that reported in \cite{Chiang2013}. We show both the backward-tracked Coma-like cluster evolution from TNG-Cluster and the \cite{Chiang2013} halo mass evolution in Figure~\ref{fig:M200_evolution_a}.

It is key to note, however, that clusters have diverse evolutionary pathways \citep{Angulo2012}. They can occasionally emerge rapidly through significant mergers of haloes from a relatively small halo at lower redshift, or from a shallower evolution from a massive halo at, for example, $z>7$. However, the candidate protoclusters at high redshift are already massive haloes, and as such, it is more appropriate to track the forward evolution of massive haloes at high-redshift down to their descendants at present day, than vice versa. This results in a shallower average halo mass evolution for massive high-redshift haloes than cluster progenitors, as has been previously shown by \cite{Lim2024} and \cite{Witten2025}. Ultimately, this highlights that the progenitors of clusters do not evolve in the same way as the descendants of massive, high-redshift haloes -- a point that has been made for both galaxy-scale and group-scale halo evolution, for example, by \cite{Behroozi2013a} and \cite{Diener2013}, respectively. As such, the most massive haloes at high redshift do not necessarily become Coma-like clusters at present day \citep{Angulo2012}. 

In order to estimate the likely descendant present-day halo mass of our protocluster candidate sample, we again utilise the TNG300 and TNG-Cluster simulations. We include in our analysis all haloes with masses $M_{200c}>10^{11}\, {\rm M_{\odot}}$ at each redshift snapshot. We utilise the \texttt{SubLink} merger trees to follow the descendant branches to establish the $z=0$ halo masses of each high-redshift halo. Given that, when analysing observations we only focus on the most massive halo within each protocluster environment (as discussed in Section~\ref{subsec:SMHM}), when numerous high-redshift haloes collapse into the same $z=0$ halo we only consider the most massive progenitor halo in this simulation analysis. With this in hand, in Figure~\ref{fig:M200_evolution_b}, we show the median present-day halo mass of $M_{200c}>10^{11}\, {\rm M_{\odot}}$ haloes in each bin of mass and redshift, over the redshift range $0<z<9$, where we have interpolated across gaps in the redshift grids of the simulations. As we are focussing on the progenitors of the largest $z=0$ haloes, if the median descendant halo mass in a given bin is less than $10^{13}\,{\rm M_\odot}$ we omit this bin from the plot. We estimate the present-day halo masses of our observed protocluster candidates by finding all haloes at the nearest redshift snapshot that have halo masses within the mass and uncertainty reported in the literature (or from our updated halo masses, reported in Table~\ref{tab:spec_sample}), and we establish the distribution of descendant present-day halo masses (discussed further in Appendix~\ref{app:dist}).

We first consider the impact of forward tracking on the halo mass evolution on the original halo masses taken from the literature (shown as transparent black points in Figure~\ref{fig:M200_evolution_b}). We model the distribution of the $z=0$ halo masses of each protocluster candidate with a log-normal distribution. We redraw the protocluster candidates from their distribution 1000 times, and for each iteration we measure the number of our protocluster candidate sample that satisfy the cluster and Coma-like cluster definitions. When using the halo masses reported in the literature, the majority of protocluster candidates remain on course to become clusters at $z=0$, as can be seen in Figure~\ref{fig:M200_evolution_b}. As a result, the number density of protoclusters remains largely unchanged, with a mild reduction in the number density of proto-Coma-clusters, albeit still $\sim 1.9$ dex offset from the expected number density (see grey data points in Figure~\ref{fig:number_density}). When we include the uncertainty as a result of cosmic variance, that was discussed in Section~\ref{sec:nominal_number_density}, in quadrature (see uncertainties in Figure~\ref{fig:number_density}), we find these results remain in strong tension with the expected number densities, at $2.7\sigma$ and $3\sigma$, respectively.

We therefore combine the improved halo mass estimates, reported in Table~\ref{tab:spec_sample}, and the more appropriate halo mass evolution, as seen in Figure~\ref{fig:M200_evolution_b}, using the same method as outlined above. When both corrections are applied, it becomes clear that none of the literature protocluster candidates are likely to have a $z=0$ halo mass of $M_{200c}>10^{15}\, {\rm M_{\odot}}$, and hence the number density of proto-Coma-clusters becomes consistent with that predicted by \cite{Tinker2008}, as shown in Figure~\ref{fig:number_density}. In addition, the number density of protoclusters significantly decreases and becomes consistent, within its $1\sigma$ uncertainty, with that expected from \cite{Tinker2008}. 

Utilising the median present-day halo mass (reported in Table~\ref{tab:spec_sample}) we find that two-thirds of the literature protocluster candidates are instead proto-groups (i.e., $10^{13}\, {\rm M_{\odot}}<M_{200c,\, z=0}<10^{14}\, {\rm M_{\odot}}$), compared to one-third that are protoclusters ($M_{200c,\, z=0}>10^{14}\, {\rm M_{\odot}}$). Notably, none of the sample of literature candidates are on course to become Coma-like clusters ($M_{200c,\, z=0}>10^{15}\, {\rm M_{\odot}}$), or indeed a lower halo mass than a galaxy group, i.e. $M_{200c,\, z=0}<10^{13}\, {\rm M_{\odot}}$, by present day.

When we account for the uncertainty in the distribution of present-day halo masses we find an expected protocluster fraction of $36\pm 9\%$ and a protogroup fraction of $54\pm 10\%$. The expected fraction of proto-Coma-clusters, and present-day haloes less massive than groups, are $4\pm 4\%$ and $9\pm 6\%$, respectively. While there are no strong proto-Coma-cluster candidates, the non-zero expected fraction underlines the broad distribution of present-day halo masses, which is discussed in more detail in Appendix~\ref{app:dist}.

Our analysis, forward tracking similar mass high-redshift haloes in simulations, allows us to account for the uncertainties in the halo mass evolution that have been previously reported by, e.g., \cite{Angulo2012}, \cite{Lim2024} and \cite{Witten2025}. These uncertainties are typically neglected when simply comparing a halo mass to the \cite{Chiang2013} halo mass evolution. However, there are potentially more reliable tracers of the present-day halo mass, for example the total mass within the Lagrangian radius. When the Lagrangian radius, $R_{\rm L}$, is considered to be the radius at which the membership probability of the present-day cluster is $50\%$, it is typically seen to be $R_{\rm L}\sim 10-15$ cMpc at $z>5$ \citep[e.g.][]{Chiang2017}. The mass within $R_{\rm L}$ is more akin to the halo masses reported in the literature that are calculated over large apertures. While we are not stating that $M_{200c}$ is the most effective diagnostic of the present-day halo mass, when observers compare a halo mass to the halo mass evolution from simulations (e.g. \citealt{Chiang2013} and \citealt{Witten2025}) it is critical that the same aperture is used as in the simulations, i.e. the virial radius.

\subsection{Do multiple protoclusters merge to form one eventual cluster?}
\label{subsec:merging}
While protoclusters at high redshift can be undergoing an inside-out growth phase \citep[i.e. dominated by star formation in the core;][]{Chiang2017}, it is plausible that within the Lagrangian radius, there are a number of massive haloes, several of which may have halo masses that place them individually on the halo mass evolutionary path of a cluster. As such, it is important to consider whether the observed protoclusters may inhabit the same eventual cluster, as previously noted by \cite{Helton2024}. The Lagrangian radius of protoclusters at $z>5$ is typically $R_{\rm L}\sim 10-15\, {\rm cMpc}$ \citep{Chiang2017,Lim2024}, and we therefore consider whether any of the protoclusters in our sample lie within coincident volumes on the sky and in redshift space. 

As a result of the relatively large redshift distances between the protoclusters reported in \cite{Terp2026} we find no coincident candidates in their sample. In contrast, we find that five of the protoclusters from the \cite{Helton2024} sample have a companion protocluster candidate from the same sample within the typical Lagrangian radius. When this occurs, we assume the halo mass to be that of the more massive candidate, and hence combine the two candidates into a single protocluster. Consequently, as shown in Figure~\ref{fig:number_density}, the number density of proto-Coma-clusters remains unchanged, however, the number density of protoclusters reduces further, falling to almost exactly the value expected from the \cite{Tinker2008} halo mass function. 

\begin{figure}
    \centering
    \includegraphics[width=1\linewidth]{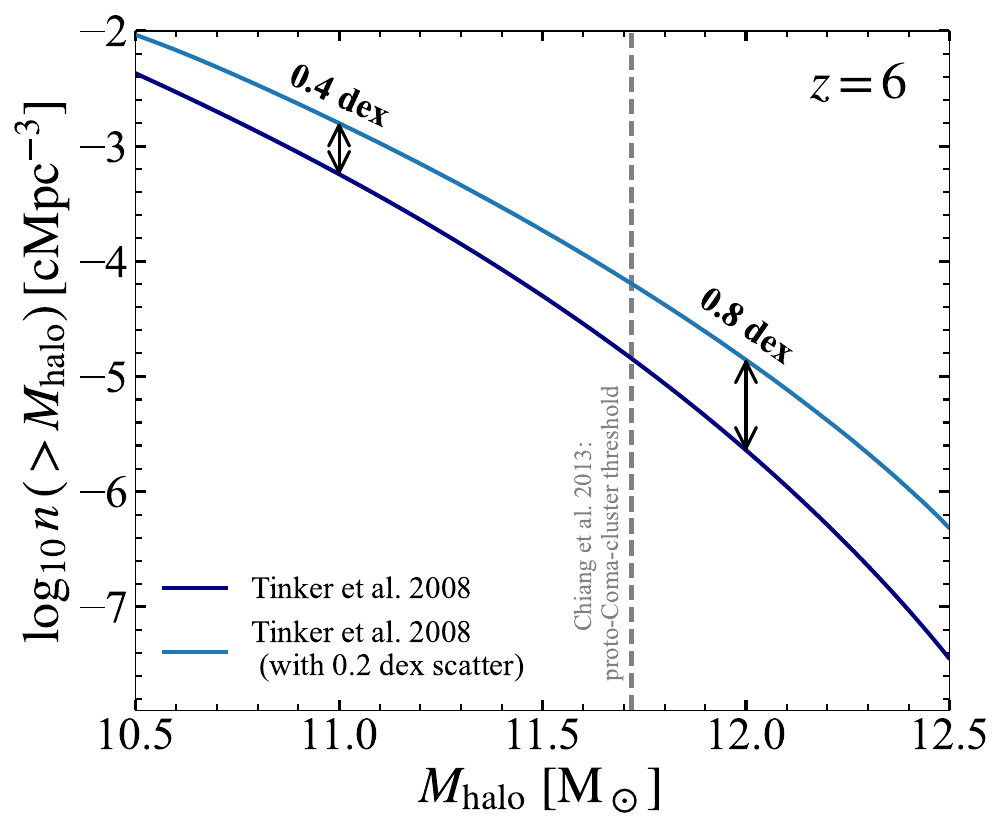}
    \caption{Halo mass function at $z=6$ from \cite{Tinker2008} (navy line), and after accounting for a 0.2 dex scatter in halo mass measurements (light blue line). The offset between the two is highlighted at $M_{\rm halo}\, [{\rm M_{\odot}}]=10^{11}$ and $10^{12}$. The threshold halo mass for a proto-Coma-cluster progenitor from \cite{Chiang2013} at $z=6$ is indicated with a grey dashed line. This highlights the impact of Eddington bias in producing a 0.65 dex bias in the observed number density of proto-Coma-clusters.}
    \label{fig:edd_bias}
\end{figure}

\subsection{How is the number density of massive haloes affected by Eddington bias?}
The halo mass function evolves rapidly with increasing halo mass (see Figure~\ref{fig:edd_bias}; \citealt{Tinker2008}). As a result, the observed number density of haloes can be significantly impacted by Eddington bias \citep{Eddington1913}. The measurement uncertainty in the value of the halo mass means that a given halo mass can scatter to higher or lower observed halo mass. When it scatters lower, given the rapidly decreasing halo mass function, it scatters into a more populated halo mass regime, making minimal difference to the distribution of halo masses. However, when the halo mass scatters upwards, it scatters into a rarely populated higher halo mass regime, resulting in a significant increase in the number density of such haloes. This produces an apparently shallower halo mass function. To assess this effect, we take the \cite{Tinker2008} $z=6$ halo mass function and redraw each halo mass from a Gaussian with a standard deviation of 0.2 dex (the typical uncertainty in the measurements presented in Table~\ref{tab:spec_sample}). The resulting bias to the halo mass function is shown in Figure~\ref{fig:edd_bias}, and results in a 0.4 dex and 0.8 dex enhancement in the number density of haloes with $M_{\rm halo}\, [{\rm M_{\odot}}]>10^{11}$ and $10^{12}$, respectively.

As we have previously noted in Section~\ref{subsec:evo}, characterising overdensities as protoclusters using the virial mass should be treated probabilistically (see Section~\ref{sec:implications} for further discussion). As a result it is challenging to make a direct comparison between how the bias of the high-redshift halo mass function affects the protocluster number density. However, given that most literature protoclusters are identified utilising a deterministic approach by comparing the observed halo mass with the \cite{Chiang2013} proto-Coma-cluster halo mass threshold, we can quantify the impact of Eddington bias. Figure~\ref{fig:edd_bias} indicates that this enhances the proto-Coma-cluster number density at $z=6$ by 0.65 dex. This is insufficient alone to resolve the nominal overabundance of proto-Coma-clusters in the literature (see Figure~\ref{fig:number_density}). However, assuming a similar bias occurs for protocluster number densities then the final number density (after correcting for the bias, the halo mass and evolution corrections, and mergers) will fall below the present-day number density of clusters. This ultimately reflects the uncertainty in protocluster selection criteria in the early Universe. We explore the potential for improving the selection of protoclusters in Section~\ref{sec:implications}.

\section{Broader implications}
\label{sec:implications}
\subsection{Improved protocluster selection}

\begin{figure}
    \centering
    \includegraphics[width=1\linewidth]{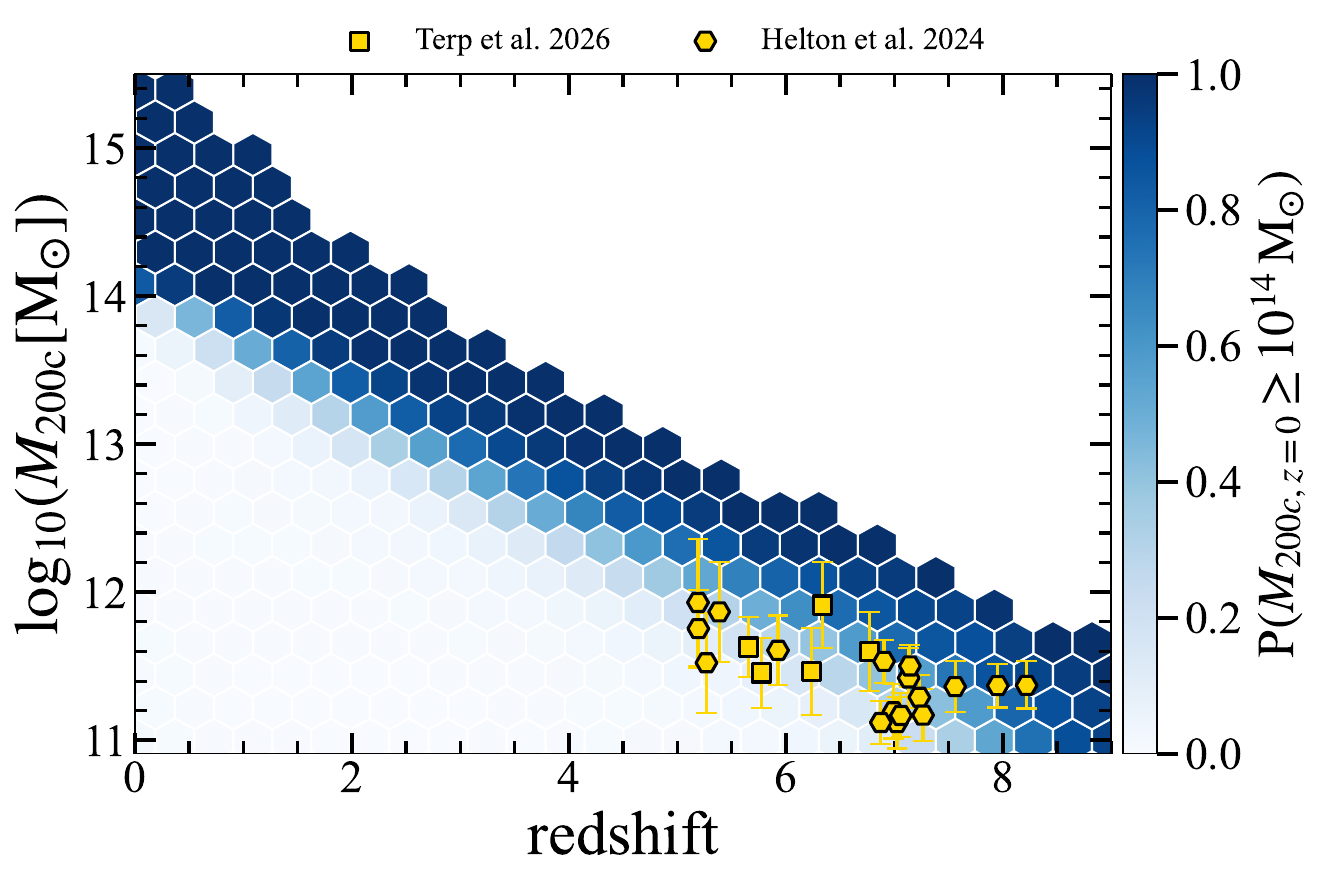}
    \caption{Probability that a simulated halo of a given mass and redshift evolves into a cluster by present day. Each hexbin is colour coded based on the fraction of simulated haloes within each bin that reach $M_{200c}\geq 10^{14}\, {\rm M_{\odot}}$ by $z=0$. Over-plotted in gold are the updated halo masses estimated in Section~\ref{subsec:SMHM}, from the \cite{Helton2024} (hexagon) and \cite{Terp2026} (square) protocluster samples. While the most massive haloes at high redshift have a high probability of becoming clusters, protocluster candidates in the literature have probabilities of $6\%-90\%$. This highlights the inherent uncertainty in a basic virial mass criterion for selecting protocluster candidates.}
    \label{fig:PC_probability}
\end{figure}

\begin{figure}
    \centering
    \includegraphics[width=1\linewidth]{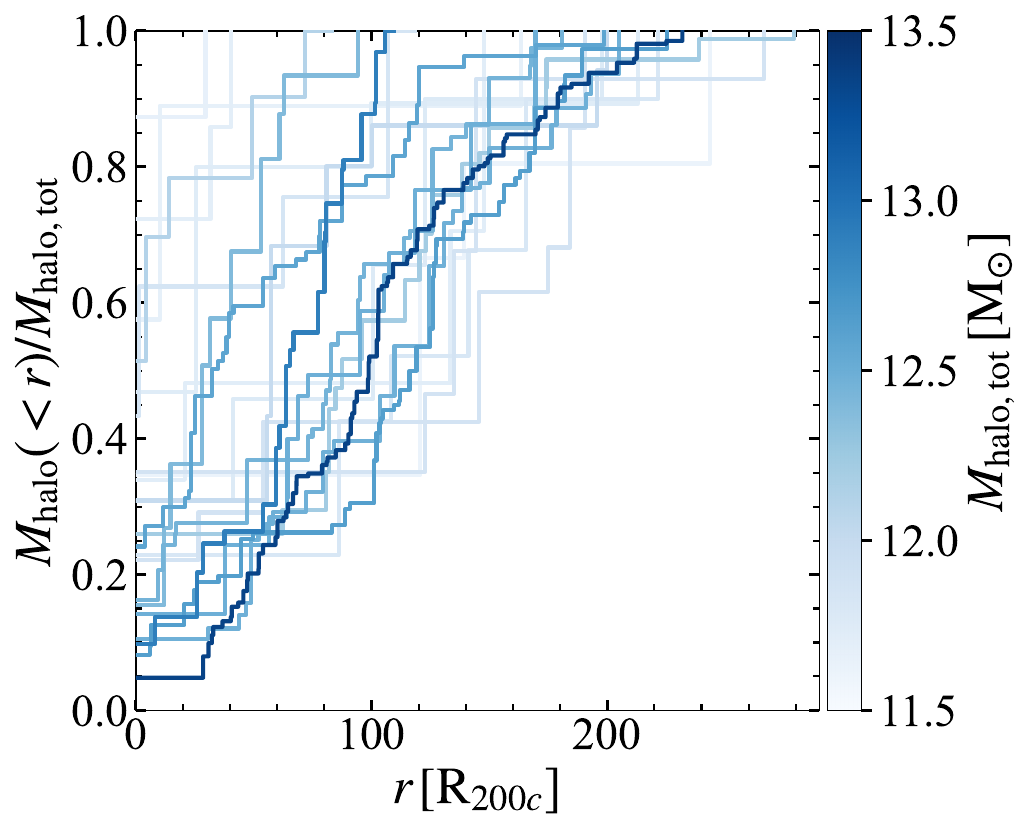}
    \caption{Fraction of the total halo mass of literature protocluster candidates enclosed within a distance from the most massive halo, in units of the virial radius (which ranges from $20-50$ pkpc). These are colour-coded based on the sum of the halo masses over the largest aperture. The range of the x-axis does not surpass the Lagrangian scales (typically of order $100\times R_{200c}$) for any of these protoclusters, but is often comparable. This highlights the significant discrepancy between the halo mass within the virial radius and within larger scales, comparable to the Lagrangian radius. While it is critical to use $M_{200c}$ for simulation comparison, considering the halo mass on Lagrangian scales may be more appropriate for identifying protoclusters.}
    \label{fig:halo_mass_apertures}
\end{figure}

As we have discussed throughout this paper, while it is critical to use consistent apertures when comparing observations to simulations, we do not explicitly prefer $M_{200c}$ as a diagnostic of present-day halo mass (see \citealt{Remus2023} for further discussion). In Figure~\ref{fig:PC_probability} we explore the probability that a simulated halo with a given mass and redshift evolves into a cluster by present day, i.e., is a protocluster. Massive haloes, comparable to those that on average become Coma-like clusters in Figure~\ref{fig:M200_evolution_b}, have a near-unity probability of becoming a cluster. However, the sample of protocluster candidates from the literature sit in a regime where the probability of being a true protocluster varies significantly. We consider the distribution of present-day descendants further in Appendix~\ref{app:dist}, and report the probability of being a protocluster in Table~\ref{tab:spec_sample}, which varies from $6\%$ to $90\%$. Such a diverse range of probabilities within a small halo mass range, combined with the discrepancies between number densities from different surveys (see Table~\ref{tab:density}), demonstrates the inaccuracy of a simple virial mass protocluster diagnostic.

In comparison to the small scale of the virial radius ($\sim 0.1$ cMpc for the sample of literature protoclusters), the Lagrangian radius ($R_{\rm L}\sim 10$ cMpc at $z>5$, \citealt{Chiang2017}) encapsulates the haloes that will eventually reside within the cluster at present day. We therefore explore the fraction of the halo mass enclosed within different apertures relative to the summation of the halo masses of all galaxies within the observed overdensity of galaxies in Figure~\ref{fig:halo_mass_apertures}, for the literature protocluster sample. We utilise the same halo mass estimate methodology as discussed in Section~\ref{subsec:SMHM}, where we split the protocluster-resident galaxies into different haloes and sum these as a function of the position from the most massive central halo of the protocluster. The largest apertures in Figure~\ref{fig:halo_mass_apertures} are up to $\sim 15$ cMpc and hence can be considered to be smaller than, or equal to, the typical Lagrangian radius. Figure~\ref{fig:halo_mass_apertures} shows that for the most massive protoclusters on Lagrangian radii scales, the mass within the virial radius contributes a very minor fraction of the total halo mass -- as low as $5\%$. For the protocluster with the largest total halo mass, the mass within the Lagrangian radius is already only a factor $\sim 4$ lower than the definition of a cluster, suggesting it is safely on course to become a cluster by $z=0$. 

At higher redshift it becomes more challenging to identify lower mass galaxies that fall below the detection limits, and as such, these potential protoclusters are often dominated by a single massive halo. These results likely indicate that a combination of $M_{200c}$ and the mass within the Lagrangian radius, accounting for the stellar mass completeness, are critical in identifying robust protocluster candidates. 

We have already shown that $64\%$ of the literature protocluster candidates actually appear to be on course to become groups, not clusters, at present day. However, while these may not ultimately represent protoclusters, there remains a strong motivation to study the galaxy overdensities identified by these studies. These extreme environments contribute more than $10\%$ of the cosmic star-formation rate density at $z>5$ \citep{Chiang2017,Lim2024}, are the first sites that will experience environmental effects at early times \citep[e.g.][]{Morishita2025,Witten2025,Witten2026} and may be key drivers of cosmic reionisation \citep[e.g.][]{Witstok2024, Scholtz2024}. 

\subsection{A larger sample of protoclusters with Roman}

In order to better understand how protoclusters evolve, a large sample of robust protocluster candidates, and indeed identification of the rare proto-Coma-cluster, is required. This ultimately requires a larger surface area survey than is possible with \jwst. Our final number density of protoclusters is in broad agreement with the halo mass function from \cite{Tinker2008}, $n\sim 10^{-5}\, {\rm cMpc^{-3}}$, and as a result, we can estimate that the expected number of protoclusters per NIRCam pointing within the redshift range $5<z<9$ ($9\times10^4 \, {\rm cMpc^3}$) is one, of course with very large scatter. In contrast, in a single {\it Nancy Grace Roman} Space Telescope ({\it Roman}; \citealt{Spergel2015}) pointing ($0.28\, {\rm degree}^2$) we expect to contain 100 protoclusters and one proto-Coma-cluster. Ultra-deep surveys, like that recommended by \cite{Bagley2026}, suggest reaching the depths of the {\it Hubble} ultra-deep field observations, with at least two {\it Roman} pointings. These depths are sufficient to identify massive galaxies in protoclusters, as has previously been demonstrated by \cite{Ishigaki2016}. We additionally note that the significantly wider area of {\it Roman} relative to \jwst\ will allow for the exploration of protoclusters across their full spatial extent ($R\sim 10\, {\rm cMpc}$). This underlines the key role that {\it Roman} can play in high-redshift protocluster science, moving from individual protocluster detections, to samples of hundreds of protoclusters, offering a similar revolution in number counts to that achieved by \jwst\ for high-redshift galaxy observations. 

\section{Conclusions}
\label{sec:conclusion}

Throughout this paper we have highlighted that the number of $z>5$ protocluster candidates reported in the literature far exceeds the expected number density of clusters at present-day. While multiple protoclusters can merge to form a single cluster by $z=0$, we have shown that this does not have a significant impact on the number density of protoclusters. The nominal observed number density of protoclusters is in $2.3\sigma$ tension with the expected number density, while the proto-Coma-cluster density is in $4\sigma$ tension.

We identify two key discrepancies when comparing observed overdensities to typical halo mass evolutionary pathways from simulations in order to diagnose them as protoclusters:
\begin{itemize}
    \item Comparisons between the halo masses summed over large observational apertures (often on megaparsec scales) to the halo masses from simulations calculated within the virial radius (typically tens of pkpc).
    \item The placement of the total halo mass onto the typical backward-traced cluster halo mass evolutionary pathway, rather than on a pathway that forward tracks massive high-redshift haloes, like those that are observationally selected.  
\end{itemize}

We have shown that {\it individually} correcting these discrepancies is incapable of sufficiently reducing the number density of protoclusters. However, in combination, these corrections do produce a number density consistent with the abundance of present-day clusters and Coma-like clusters, within nearly $1\sigma$ uncertainties. After considering the impact of coincident protoclusters merging, cosmic variance and Eddington bias, we can entirely alleviate any tension with the expected number density. 

These corrections ultimately imply that $\sim 64\%$ of reported protoclusters are in fact likely to become galaxy groups ($M_{200c}< 10^{14}\, {\rm M_{\odot}}$) by present day. Moreover, none of the observed candidates represents a robust progenitor of a Coma-like cluster. Regardless, these environments remain some of the most extreme overdensities in the early Universe, where the first environmental effects are expected to occur, and hence they warrant the attention that the community supplies them. Identification of broader samples of protoclusters and indeed proto-Coma-clusters is feasible in the foreseeable future with the vast volumes that will be probed by the {\it Roman} space telescope.

In summary, our work indicates that the evolutionary pathways of high-redshift massive haloes show a variety of different present-day masses (see Appendix~\ref{app:dist}), and therefore, identifying a protocluster from $M_{200c}$ alone is unreliable. Instead, we propose that the halo mass within the Lagrangian radius ($\sim 10$ cMpc; akin to the scales over which observers typically calculate the halo masses of galaxy overdensities) may provide a more robust diagnostic of the present-day halo mass. However, when observers choose to utilise the $M_{200c}$ halo mass evolutionary pathways from simulations to diagnose a protocluster, it is vital to ensure the apertures used in their observations match those from simulations.

\begin{acknowledgements}
The work presented in this paper is based on observations made with the NASA/ESA/CSA James Webb Space Telescope. The data were obtained from the Mikulski Archive for Space Telescopes at the Space Telescope Science Institute, which is operated by the Association of Universities for Research in Astronomy, Inc., under NASA contract NAS 5-03127 for JWST. 

This work has received funding from the Swiss State Secretariat for Education, Research and Innovation (SERI) under contract number MB22.00072, as well as from the Swiss National Science Foundation (SNSF) through project grant 200020\_207349. JSB acknowledges support from the Simons' Collaboration on Learning the Universe and a Leverhulme Trust Early Career Fellowship.
The Cosmic Dawn Center (DAWN) is funded by the Danish National Research Foundation under grant DNRF140. JMH acknowledges support from JWST Program 8544.
KEH acknowledges support from the Independent Research Fund Denmark (DFF) under grant 5251-00009B.
SL acknowledges support by the Science and Technology Facilities Council (STFC) and by the UKRI Frontier Research grant RISEandFALL. 
\end{acknowledgements}

\bibliographystyle{aa}
\bibliography{references} 

\begin{appendix}
\section{Literature protocluster sample}
In this work we have taken a number of candidate protoclusters reported by literature spectroscopic searches over large volumes from \cite{Helton2024} and \cite{Terp2026}, reported in Table~\ref{tab:spec_sample}. We additionally utilise the photometric sample from \cite{Li2025} to evidence the overabundance of protoclusters in a photometric sample in Table~\ref{tab:density}. We report the photometrically-identified protocluster candidates from \cite{Li2025} in Table~\ref{tab:phot_sample}.

\begin{table*}[]
    \centering
    \caption{Spectroscopic sample of protocluster candidates reported in the literature.}
    \label{tab:spec_sample}
    \begin{tabular}{lcccccc}
    \hline
    Name & $\left<z \right>$ & $N_{\rm gal}$ & ${\rm log}_{10}(M_{\rm halo,\, lit.}\,[{\rm M_{\odot}}])$ & ${\rm log}_{10}(M_{200c}\,[{\rm M_{\odot}}])$ & P$(M_{200c, z=0}\geq 10^{14}\, {\rm M_{\odot}})$ & ref. \\ 
    \hline
    A2744-PC-z5p7 & 5.66 & 34 & $12.05^{+0.27}_{-0.26}$ &$11.63\pm0.20$ & 0.17 & [1]\\
    A2744-PC-z5p8 & 5.77 & 59 & $12.06^{+0.26}_{-0.19}$ &$11.45\pm0.24$& 0.11 & [1] \\
    A2744-PC-z6p2 & 6.24 & 33 & $11.95^{+0.24}_{-0.18}$ &$11.46\pm0.30$&0.21 & [1]\\
    A2744-PC-z6p3 & 6.34 & 43 & $12.54^{+0.41}_{-0.21}$ &$11.91\pm0.29$&0.61 & [1] \\
    A2744-PC-z6p8 & 6.77 & 21 & $11.84^{+0.18}_{-0.34}$ &$11.60\pm0.27$&0.50 & [1]\\
    JADES-GN-OD-5.191 & 5.19 & 103 & $13.44 \pm 0.04$ & $11.92 \pm 0.43$ &0.21 & [2] \\
    JADES-GN-OD-5.194 & 5.19 & 8 & $12.25 \pm 0.16$ & $11.76 \pm 0.26$ &0.14 & [2] \\
    JADES-GN-OD-5.269 & 5.27 & 14 & $12.38 \pm 0.07$ & $11.53 \pm 0.34$ &0.06 & [2] \\
    JADES-GS-OD-5.386 & 5.39 & 39 & $13.04 \pm 0.04$ & $11.87 \pm 0.34$ &0.24 & [2] \\
    JADES-GS-OD-5.928 & 5.93 & 14 & $12.50 \pm 0.06$ & $11.61 \pm 0.24$ &0.22 & [2] \\
    JADES-GS-OD-6.876 & 6.88 & 4 & $11.46 \pm 0.12$ & $11.12 \pm 0.14$ &0.13 & [2] \\
    JADES-GS-OD-6.906 & 6.91 & 4 & $11.81 \pm 0.15$ & $11.53 \pm 0.15$ &0.61 & [2] \\
    JADES-GN-OD-6.991 & 6.99 & 6 & $11.66 \pm 0.11$ & $11.20 \pm 0.19$ &0.28 & [2] \\
    JADES-GN-OD-7.025 & 7.03 & 6 & $11.61 \pm 0.10$ & $11.12 \pm 0.18$ &0.23 & [2] \\
    JADES-GS-OD-7.061 & 7.06 & 5 & $11.63 \pm 0.13$ & $11.17 \pm 0.15$ &0.23 & [2] \\
    JADES-GN-OD-7.133 & 7.13 & 5 & $11.85 \pm 0.15$ & $11.42 \pm 0.20$ &0.52 & [2] \\
    JADES-GN-OD-7.144 & 7.14 & 4 & $11.83 \pm 0.20$ & $11.50 \pm 0.14$ &0.62 & [2] \\
    JADES-GS-OD-7.231 & 7.23 & 5 & $11.73 \pm 0.14$ & $11.29 \pm 0.15$ &0.44 & [2] \\
    JADES-GS-OD-7.265 & 7.27 & 7 & $11.68 \pm 0.11$ & $11.17 \pm 0.18$ &0.37 & [2] \\
    JADES-GS-OD-7.561 & 7.56 & 7 & $11.91 \pm 0.12$ & $11.36 \pm 0.17$ &0.70 & [2] \\
    JADES-GS-OD-7.954 & 7.95 & 4 & $11.61 \pm 0.15$ & $11.37 \pm 0.15$ &0.87 & [2] \\
    JADES-GS-OD-8.220 & 8.22 & 6 & $11.79 \pm 0.14$ & $11.37 \pm 0.16$ &0.90 & [2] \\
    \hline
    \end{tabular}
    \\
    References: [1] \cite{Terp2026}, [2] \cite{Helton2024}.
\end{table*}

\begin{table*}[]
    \centering
    \caption{Photometric sample of protocluster candidates reported in \cite{Li2025}.}
    \label{tab:phot_sample}
    \begin{tabular}{lcccc}
    \hline
    Name & $\left<z \right>$ & $N_{\rm gal}$ & ${\rm log}_{10}(M_{\rm halo,\, lit.}\,[{\rm M_{\odot}}])$ \\ 
    \hline
    CEERS-ID-1 & 5.16 & 32 & $12.27^{+0.52}_{-0.29}$ \\
    CEERS-ID-2 & 5.19 & 9 & $12.12^{+0.40}_{-0.24}$ \\        
    CEERS-ID-3 & 5.40 & 6 & $11.63^{+0.10}_{-0.16}$  \\
    CEERS-ID-4 & 5.40 & 20 & $13.57^{+0.83}_{-1.07}$  \\
    CEERS-ID-5 & 5.42 & 5 & $11.54^{+0.13}_{-0.10}$  \\
    CEERS-ID-6 & 5.47 & 5 & $11.45^{+0.06}_{-0.12}$  \\
    CEERS-ID-7 & 5.79 & 10 & $11.61^{+0.15}_{-0.12}$  \\
    CEERS-ID-8 & 5.93 & 6 & $12.65^{+0.55}_{-0.58}$  \\
    CEERS-ID-9 & 6.29 & 25 & $11.52^{+0.17}_{-0.13}$  \\
    CEERS-ID-10 & 6.76 & 12 & $11.28^{+0.11}_{-0.12}$ \\
    NEP-ID-1 & 5.34 & 14 & $11.88^{+0.29}_{-0.20}$  \\
    NEP-ID-2 & 5.50 & 37 & $12.40^{+0.69}_{-0.48}$ \\
    NEP-ID-3 & 5.57 & 5 & $11.17^{+0.10}_{-0.10}$ \\
    NEP-ID-4 & 5.74 & 29 & $12.46^{+0.50}_{-0.45}$\\
    NEP-ID-5 & 6.08 & 41 & $13.92^{+0.48}_{-0.63}$\\
    NEP-ID-6 & 6.29 & 12 & $11.61^{+0.27}_{-0.16}$  \\
    NEP-ID-7 & 6.41 & 37 & $11.94^{+0.51}_{-0.27}$ \\
    NEP-ID-8 & 6.69 & 13 & $11.83^{+0.46}_{-0.23}$\\
    NEP-ID-9 & 6.75 & 5 & $11.29^{+0.14}_{-0.11}$ \\
    NEP-ID-10 & 6.90 & 20 & $11.70^{+0.78}_{-0.34}$ \\
    \hline
    \end{tabular}
\end{table*}

\section{Distribution of descendant halo masses}
\label{app:dist}
\begin{figure*}
    \centering
    \includegraphics[width=0.58\linewidth]{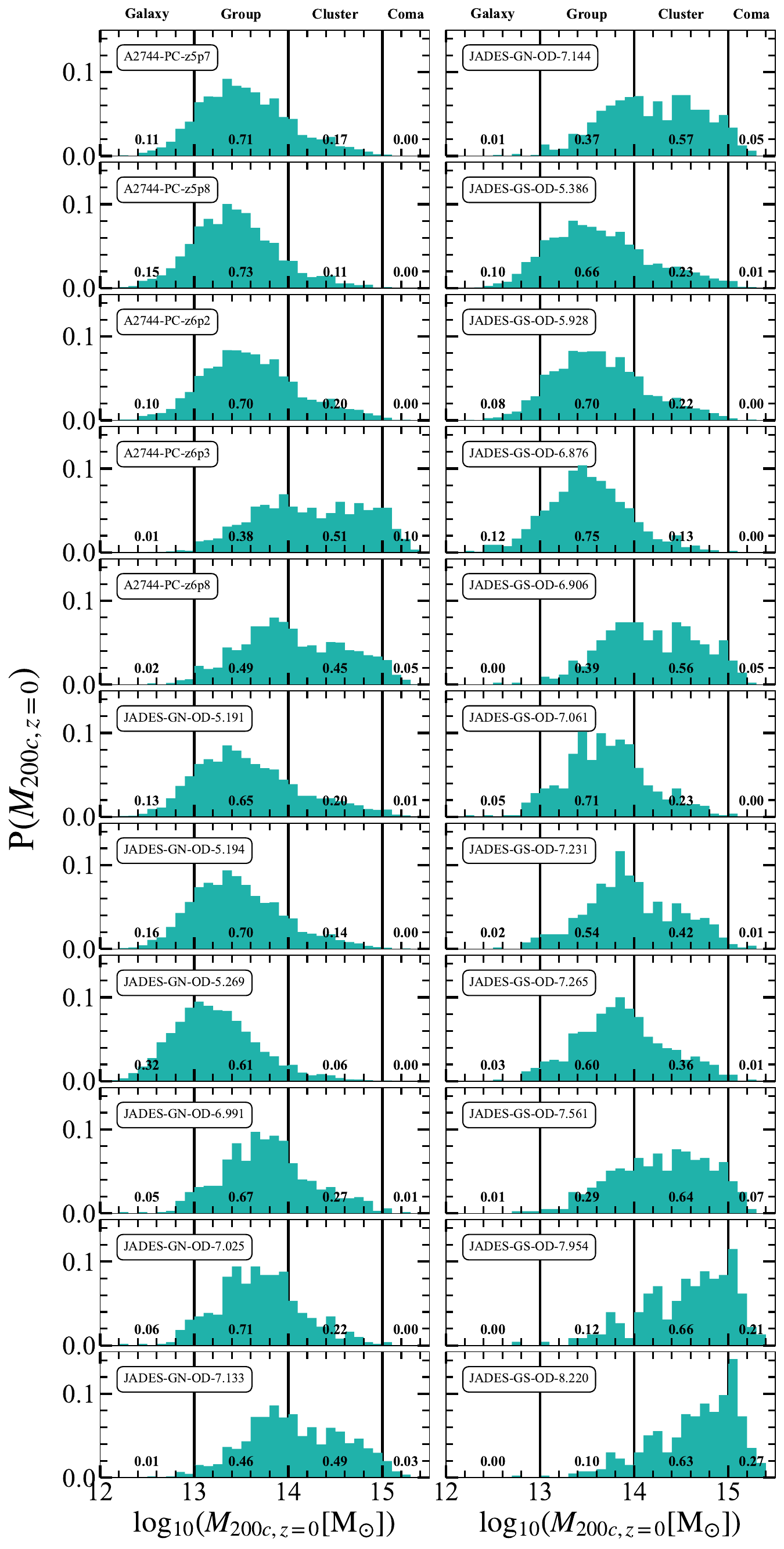}
    \caption{Probability distribution of present-day halo masses of simulated haloes matched to the updated halo mass and redshift of the sample of literature protocluster candidates. Each panel represents the distribution for each protocluster, indicated by the name in the top-left of the panel. The panel is split into four mass regimes: galaxy-like haloes, groups, clusters and Coma-like clusters. The probability of the protocluster evolving into each of these mass regimes is indicated with text at the bottom of each panel. Two-thirds of the sample are best characterised as proto-groups, while the other third appear to be protoclusters, and none of the sample represent robust proto-Coma-clusters. The broad distribution of present-day halo masses evidences the uncertainty of a basic virial mass diagnostic for protocluster candidates.}
    \label{fig:halo_mass_dist}
\end{figure*}

To understand the distribution of possible descendant halo masses at present day for the literature sample of protocluster candidates, and indeed their probability of becoming a cluster at present day, we once again employ the TNG300 and TNG-Cluster simulations. Following the same method as Section~\ref{sec:resolution} we identify simulated haloes at the same redshift, and within the halo mass uncertainty, as the observed protocluster candidates in Table~\ref{tab:spec_sample}. We then show the distribution of their present-day descendants in Figure~\ref{fig:halo_mass_dist}, and utilising this we estimate the probability that each observed protocluster candidates is indeed a protocluster, and report this in Table~\ref{tab:spec_sample}.

While around one-third of the candidates have a greater than $50\%$ probability of becoming clusters at present day, none of the candidates have a strong probability of becoming a Coma-like cluster. All of the candidates have a broad distribution of present-day halo masses, evidencing the uncertainty of a purely virial mass based selection of protocluster candidates.

\end{appendix}

\end{document}